# Spatiotemporal flat optics for terabit-per-second single-channel data transmission

Wen-Jing Liu[1,#], Dong Zhao[1,#,*], Heng-Yi Wang[1,#], Jun He[1], Fang-Wen Sun[1], Ye Tian[2,3,*], Kun Huang[1,4,*]

[1]Department of Optics and Optical Engineering, University of Science and Technology of China, Hefei, 230026, Anhui, China

[2]State Key Laboratory of Ultra-intense Laser Science and Technology, Shanghai Institute of Optics and Fine Mechanics (SIOM), Chinese Academy of Sciences (CAS), Shanghai 201800, China

[3]Center of Materials Science and Optoelectronics Engineering, University of Chinese Academy of Sciences, Beijing 100049, China

[4]State Key Laboratory of Opto-Electronic Information Acquisition and Protection Technology, Anhui University, Hefei, 230601, Anhui, China

**Abstract**

**Exponential growth in global data traffic demands ever-increasing transmission rates—a pursuit fundamentally constrained by the physical limitations of digital-to-analog converters (DACs). Existing strategies to overcome this bottleneck, such as multi-DAC arrays and optical time-division multiplexing, inevitably introduce system complexity and coordination overhead. Here we demonstrate an all-optical spatiotemporal transmitter that generates controllable high-repetition information-carrying femtosecond pulses at the focus of a phase-modulated planar diffractive lens (PDL) through optical-path-induced spatial-to-temporal conversion. Each pulse serves as an information bit, encoding binary data via on-axis focal intensity states corresponding to "0" and "1", achieved by switching between topological and constant phase modulations. High experimental orthogonality between arbitrary bits enables nearly error-free transmission of 15×15-pixel grayscale (8-bit coding) and colour (9-bit coding) images at a record-high single-channel rate of ~3 terabits per second (Tbit/s). Free from electronic and coordination bottlenecks, this all-optical transmitter establishes a scalable high-speed single-channel pathway toward ultrahigh-capacity optical communication.**

**Introduction**

The rapid advancement of artificial intelligence, cloud computing, global connectivity, and the Internet of Things (IoT) is driving an unprecedented demand for high-speed data transmission[1-4]. To support this growth, next-generation Ethernet standards are targeting aggregate rates of 1.6 Tbit/s and 3.2 Tbit/s[5]. Within data centre networks, intensity-modulation direct-detection (IMDD) method remains the dominant solution owing to its simplicity and energy efficiency[6]. Realizing such capacities on a single channel, however, presents a far more formidable challenge, defining the pivotal frontier for IMDD technology.

The capacity of a single channel is fundamentally governed by signal bandwidth and spectral efficiency (SE)[7]. Under fixed SE constraints, the achievable data rate scales linearly with bandwidth, rendering bandwidth expansion the most direct route to higher transmission speeds. This route, however, confronts a foundational bottleneck in the transmitter's DAC[8-13]. State-of-the-art current-steering DACs, widely adopted in high-speed optical communications[13-16], exhibit intrinsic 3-dB bandwidths of tens of gigahertz for a single device[13]. This limit is rooted in semiconductor physics such as low carrier mobility, limited drive voltage, parasitic capacitance effects and long channel length[16-21]. Simultaneously, the pursuit of higher SE through increased DAC resolution is self-limiting: it exacerbates noise, timing jitter and nonlinearity[22-25], while precipitating a strict bandwidth-resolution trade-off that confines practical operation to ~8 bits[23,26-31]. Thus, the DAC imposes this dual constraint on bandwidth and SE, which defines the current ceiling for single-channel rates. Although multi-DAC approaches can partially alleviate bandwidth constraints, synchronization errors and channel mismatches inevitably degrade signal quality[32-35]. As an alternative pathway, optical time-division multiplexing (OTDM)[36] involves the optical synthesis of an ultrahigh-speed optical pulse train via interleaving time-delayed optical pulses (Fig. 1a). To encode this optical carrier, the high-speed data stream is demultiplexed into multiple parallel low-speed electrical channels, which are then converted by synchronized DACs to drive the modulator, thereby operating within a relaxed per-channel bandwidth requirement. However, the required coordination among multiple electronic and optical components substantially increases system complexity[37-40]. These limitations motivate the development of new approaches that circumvent electronic bandwidth constraints.

Alternatively, all-optical information processing has emerged as a promising method toward low power consumption and high operational speed[41]. Recent advances in spatiotemporal manipulation of optical pulses provide additional degrees of freedom for controlling signals[42-45], facilitating data transmission. Yet, harnessing this potential requires precise synthesis of tailored spatiotemporal fields—a task that remains a fundamental challenge, especially for ultrashort pulses. The mainstream approach relies on Fourier-transform-based pulse shaping via an optical 4-$f$ setup[46,47], which, with a spatial light modulator (SLM) at the spectral plane, can generate complex waveforms such as ultrafast vortex pulse strings[45] and vortex combs[48]. However, its angular field between the grating and lens or the finite pixel size of the SLM restricts the spatial extent of encoded pulses, and advanced encoding schemes like vortex topological multiplexing are incompatible with standard optical communication architectures. Consequently, developing a scalable, single-channel system capable of terahertz-rate operation while remaining compatible with existing frameworks remains a considerable challenge.

Here, we present a compact transmission architecture that encodes data into customizable spatiotemporal wave packets, offering scalability and tunability through concentric phase engineering. The coding process is converted into radially varying phase profiles, exploiting the differential time delays of a PDL upon focusing. Information is recovered via direct time-resolved intensity detection at the focus of the planar lens and the data rate is independently tuned by adjusting each zone's temporal delay via optical path. To validate it, we experimentally demonstrate grayscale (8-bit coding) and colour (9-bit coding) image transmission at the rates of 2.86 and 3.33 terabits/s, respectively. Our approach supports continuous operation with more coding bits by scaling the number of phase-coding zones. By unifying optical spatial patterning with ultrafast temporal control, this work establishes a structural all-optical paradigm that intrinsically bypasses electronic bandwidth bottlenecks. Its scalable and tunable features deliver unprecedented single-channel capacity within compatible systems, paving the way to future high-throughput networks.

**Data transmission via spatiotemporal flat optics**

Unlike conventional OTDM systems—which rely on multiple electronic and optical components (Fig. 1a)—the adoption of spatiotemporal wave packets (STWPs) enables a more integrated and compact architecture for ultrafast data transmission. As depicted in Fig. 1b, an incident pulse ($\lambda$ = 1.027 μm, duration 200 fs) is modulated by a encoded phase and then focused by a PDL (where a binary-phase Fresnel plate is used here), forming the proposed transmitter. This produces a time-evolving optical intensity distribution at the focal plane, termed an STWP. At specific time slices, the STWP profile exhibits either a central dark hole (zero on-axis intensity) representing binary "0", or a focused spot (nonzero on-axis intensity) representing binary "1". Thus, STWPs serve as data carriers, enabling binary data retrieval through time-gated analysis of the intensity profiles at the PDL's focal plane.

To ensure correct data transfer, the temporal profile of the STWP must be precisely controlled according to the input binary code. During encoding, a programmable phase mask—the encoded phase—is partitioned into $N$ concentric annular zones of equal duration. Each zone corresponds to one bit of an $N$-bit binary sequence $b_1b_2b_3...b_n$ (with $b_n \in \{0, 1\}$, $n = 1, 2, ..., N$). For the $n$-th zone, the phase modulation is given by $U_n = \exp(il_n\theta_0)$, where $\theta_0$ denotes the azimuthal angle, and the topological charge $l_n$ is the encoding parameter. For a bit "1", $l_n = 0$ creates an on-axis hotspot after optical focusing by the diffractive planar lens; for a bit "0", $l_n=6$ generates a null on-axis intensity. Thus, the modulation factor applied to each zone follows:

$$U_n = \begin{cases} \exp(i6\theta_0), & b_n = 0 \\ 1, & b_n = 1 \end{cases} \quad (1)$$

Figure 1c illustrates the principle of this encoding scheme. The encoded phase is segmented into eight annular zones ($P_n$, with $n = 1$ to 8), each with a temporal duration of $\Delta t = 350$ fs. To ensure precise phase mapping, the structure of this encoded phase must be spatially aligned with the geometry of the PDL. Accordingly, the outer radius $r_n$ of each zone is determined by the focal length of the predefined PDL:

$$r_n = \sqrt{D_n^2 - f^2} = \sqrt{(f + cn\Delta t)^2 - f^2}, \qquad n = 1, 2, \cdots, N \quad (2)$$

where $c$ denotes the speed of light in vacuum, $N=8$ for this coding sequence and $f$ is the PDL's focal length. Equation (2) provides the theoretical link between the spatial position $r_n$ on the PDL and the time delay of the on-axis pulse at the focal plane, forming the fundamental basis of spatial-to-temporal conversion.

To encode this exemplified sequence "01010101" (Fig. 1c), we set $U_n = \exp(i6\theta_0)$ for odd-numbered zones and $U_n = 1$ for even-numbered zones (bit indices from left to right correspond to zones $P_1$–$P_8$; Fig. 1c). The encoded phases are then imprinted onto the PDL by the input pulses. Owing to distinct optical path differences ($D_1$ to $D_8$) from zones $P_1$–$P_8$ to the focal point, the encoded pulses arrive at the focal region with proportional delays. When the pulse from the PDL's geometric center (propagating along the optical axis) reaches the focus, the pulse from the outer edge of zone $P_1$ is delayed by 350 fs, and pulses from the remaining zones arrive at correspondingly later times. This creates a chain of data-carrying sub-pulses without the need for electronic DACs. A mathematical description of this transmission principle is provided in the Methods.

Figure 1d presents time-resolved intensity distributions simulated at the peak-intensity times of each created sub-pulse. Notably, the hotspots appear at the focus at times $t_2$, $t_4$, $t_6$, and $t_8$, arising from non-vortical phase encoded onto zones $P_2$, $P_4$, $P_6$, and $P_8$. Conversely, the focused light with a vortex phase ($l_n = 6$) yields annular fields with dark centers, resulting in diminished energy density within the corresponding detection regions, as indicated by magenta circles at $t_1$, $t_3$, $t_5$, and $t_7$ in Fig. 1d. Thus, by correlating the coding sequence with the spatiotemporal intensity directly, deterministic phase-to-intensity mapping is achieved, enabling the input information to be decoded through time-resolved analysis of the intensity dynamics in the detection region.

**Experimental verification of spatiotemporal encoding/decoding**

The experimental configuration of data transmission using spatiotemporal optics is illustrated schematically in Fig. 2a. A Mach–Zehnder interferometer is employed to resolve the spatiotemporal profile, as the ultrafast dynamics lie beyond the temporal resolution of conventional detectors (e.g., CCD or CMOS sensors). Femtosecond pulses from a mode-locked laser are first filtered by a pinhole and then separated by a beam splitter ($BS_1$) into signal and reference arms. In the signal arm, a variable optical delay line provides controlled temporal offsets for subsequent interference measurements. The beam is spatially expanded by a home-built Galilean beam expander (lenses $L_1$ and $L_2$) and then illuminates a spatial light modulator (SLM) loaded with a zone-segmented phase pattern for spatiotemporal pulse shaping. To encode the sequence "10000000", only the first zone carries a uniform constant phase, whereas the remaining zones are encoded with a vortex phase (Fig. 2b). A 4-$f$ imaging system (lenses $L_3$ and $L_4$) maps the SLM-modulated phase onto a binary-phase PDL (Fig. 2c). At the focal plane of the PDL, a train of spatiotemporal sub-pulses is created and imaged onto a camera by an objective. To enhance the temporal resolution, a pair of chirped mirrors is used in the reference arm to narrow the pulse width through dispersion compensation. As the optical delay line is scanned, time-resolved interference patterns between the signal and reference pulses are recorded by the camera (Fig. 2d). Retrieving the spatiotemporal intensity profiles from these patterns yields the distributions shown in Fig. 2e: a focused spot at $t=175$ fs and ring-shaped profiles at the other time slots from $t=525$ fs to $t=2625$ fs, with an interval of $\Delta t=350$ fs. The retrieval method is detailed in Supplementary Section 1. Thus, by programming the phase pattern on the SLM according to the input code, the spatiotemporal distributions of the wave packet can be characterized experimentally for optical decoding, confirming the feasibility of this approach for data transmission.

**Encoding/Decoding Orthogonality**

To test the efficacy and evaluate the temporal orthogonality of our space-time encoding, we first transmitted an 8-bit orthogonal matrix via STWPs, as illustrated in Fig. 3. Numerical simulations based on time-harmonic Rayleigh-Sommerfeld diffraction theory[49,50] (see Methods) reveal distinct focal-region intensity profiles for different codes (Fig. 3a). For instance, activating only the focusing phase produces consecutive focal spots (code "11111111"), whereas applying a uniform topological charge of $l$=6 to all zones (code "00000000", denoted $C_0$) yields no detectable focal energy within the target region, establishing a well-defined null state. Crucially, for codes $C_1$–$C_8$, a single "1" bit shifts consecutively across eight temporal positions, translating the focal spot linearly in time across consecutive 350-fs windows (e.g., 0–350 fs for $C_1$ to 2450–2800 fs for $C_8$). This one-to-one mapping between bit position and arrival time demonstrates the encoding concept and reveals a clear trend toward temporal orthogonality. Experimental results (Fig. 3b) faithfully reproduce these simulated temporal shifts, directly confirming this spatiotemporal mapping principle.

Decoding orthogonality requires quantifying the temporal intensity evolution within a defined detection region. For the focal state "11111111", the normalized total intensity initially surges and then gradually decays (Fig. 3c). The agreement between simulation and experiment validates the operating mechanism and establishes a benchmark to normalize the detected intensity in subsequent decoding. Figure 3d compares the total intensity for codes $C_0$–$C_8$ within the same region, normalized to the focal-state intensity at the corresponding time. Whereas $C_0$ shows no distinct peaks across the 0–2800 fs, the sequences from $C_1$ to $C_8$ display well-defined peaks at 350-fs intervals, with peak positions shifting linearly as the "1" bit moves towards the ending of the coding sequence—a trend corroborated by experiment (dashed lines, Fig. 3d). Occasional normalized intensities exceeding unity arise from temporal measurement jitter. Applying an intensity threshold of 0.3 to these normalized experimental profiles yields discrete binary signals (Fig. 3e), where a high level corresponds to "1", and a low level to "0". For codes $C_1$-$C_8$, each "1" bit generates a high digital level strictly confined to its designated 350-fs window, while all "0" bits remain low. This spatiotemporal confinement directly confirms the temporal orthogonality of these encoded bits.

Such orthogonality renders continuous detection unnecessary in experiment. Accordingly, we adopt an optimal sparse-sampling strategy that acquires data solely at the midpoint of each 350-fs temporal window, drastically reducing the acquisition time in realistic data transmission. The resultant confusion matrix retains clear inter-bit orthogonality among eight temporal sampling nodes (Fig. 3f). Critically, this scheme of allocating one bit per time interval directly yields a space-time optical field transmission speed of 2.86 Tbit/s.

**8-bit coding for grayscale image transmission**

To evaluate the data-transmission performance, we transmitted a 15×15-pixel grayscale image (Fig. 4). Each pixel's 256 gray levels were encoded into a unique 8-bit binary sequence (e.g., the gray level 229 as "11100101"; Fig. 4a). Each 8-bit code was then mapped sequentially onto concentric zones from $P_1$ to $P_8$, where the bit index from left to right corresponds to the zone index. Zones with a corresponding bit of "1" received a topological charge of 0, and those for a "0" bit received charge 6. After transmission, the spatiotemporal field was sampled at eight discrete time nodes. The resulting intensity distributions near the focus (Fig. 4b) reveal distinct focal spots precisely at the temporal windows assigned to "1" bits (e.g., at 175 fs, 525 fs). Quantifying the normalized total intensity within each detection region (black circles in Fig. 4b) yields the received intensity profile (blue histogram, Fig. 4c). All signals corresponding to "1" bits consistently exceed the 0.3 decoding threshold (red line), whereas all "0" bits remain below it. Chronological decoding of this profile perfectly reconstructs the original 8-bit sequence "11100101" (orange bars) without error. Extending this decoding protocol to all 225 pixels successfully recovers the full image (right panel of Fig. 1), achieving a bit error rate (BER) of zero. This error-free transmission of a multi-gray-level image validates the system's functionality for handling complex data streams—a capability directly enabled by the engineered temporal orthogonality between bits.

Signal quality was systematically evaluated by first quantifying the intensity distributions for logic "1" and "0" at each bit position (Fig. 4d). The average intensity for a logic "1" (blue bars) consistently exceeded 0.6, with the minimum value remaining above 0.3. Conversely, the average for a logic "0" (orange bars) remained below 0.1, with the maximum never exceeding 0.3. This well-defined separation establishes 0.3 as a robust discrimination threshold. To assess transmission accuracy across the entire gray-level range, we then calculated the root-mean-square error (RMSE) between the received intensities and ideal digital values for all 256 gray levels (Fig. 4e). The RMSE is defined as $RMSE = \sqrt{\frac{1}{N}\sum_{n=1}^{N}[I_{Exp}(n) - I_{Sent}(n)]^2}$, where $N = 8$, and $I_{Exp}$ is the normalized experimental intensity, and $I_{Sent}$ is the ideal digital intensity. The resulting RMSE values for all gray levels fell mostly within the narrow range of 0.1 to 0.3. This uniformly low deviation confirms high-fidelity signal transmission across the entire dynamic range, exhibiting good robustness for data transmission.

**9-bit coding for color image transmission**

To achieve higher data rates, we scaled the system to 9-bit operation by reducing the temporal duration to $\Delta t$=300 fs. This preserves the underlying physical principles while partitioning the encoding phase into nine concentric zones, each introducing 300-fs delay. Orthogonal-matrix transmission confirmed that inter-bit orthogonality is retained (Supplementary Section 3). With this design, we achieved a single-channel transmission rate of 3.33 Tbit/s—a 16.7% improvement over the 8-bit architecture.

For full-colour image transmission, each pixel (see Fig. 5a) was encoded via its RGB channels, with each channel assigned a 3-bit value (9 bits per pixel). The 0–255 range of each channel was uniformly quantized into eight levels (step size: 36). The resulting composite 9-bit codes dictated the phase pattern across nine concentric zones (Fig. 5a, right), following the encoding principles described above. Five transmission trials yielded the received images in Fig. 5b. The measured BER ranged from 0% (Fig. 5b, i, v) to a maximum of 0.148% (Fig. 5b, iii, iv). Despite occasinal bit errors in these trials, the received images remain visually indistinguishable from the originals—minimal chromatic shifts in erroneous pixels are highlighted in light-blue boxes. Even in this worst-case scenario, only three pixels exhibited errors, each by a single bit. These results establish the scheme's robustness and high fidelity for colour image transmission, exhibiting the potential in achieving error-free performance under optimal conditions.

System performance was further evaluated by calculating the RMSE between sent and received intensities across all five trials (Fig. 5c). The RMSE values were uniformly low (0.1–0.4), indicating no systematic bias in color transmission. Analysis of the normalized intensity distributions for each bit position pinpointed the error source (Fig. 5d). Although the mean intensities for "1" (blue) and "0" (pink) were generally separable by a discrimination threshold, a critical overlap occurred specifically at the eighth bit ($B_8$), where this threshold intersected the intensity ranges of both states. BER statistics of each bit position verified this exclusive localization: all errors originated from $B_8$, whereas all other bits were transmitted error−free (Fig. 5e). To investigate the underlying cause, we calculated inter-logic-state crosstalk using data from Supplementary Fig. 3f (Fig. 5f). Crosstalk for the *n*-th bit is defined as $Crosstalk(n) = I_{nn}/\max(I_{nj})_{j \neq n}$, where *n* and *j* is valued between 1 and 9, *j* denotes the *j*-th temporal sampling node and *I* the normalized total intensity. Crosstalk peaked at $B_8$ with a value of 0.33, implying increased sensitivity to environmental noise (e.g., jitter). This crosstalk-induced vulnerability directly accounts for the observed bit errors—the elevated crosstalk at $B_8$ precisely aligns with its dominant contribution to the system's BER (Fig. 5 e, f). In our current experimental setup, the high crosstalk at $B_8$ originates from the narrow (<350 fs, see Fig. 3e and Supplementary Fig. 3e) time interval after binary digitalization of the normalized intensity. This engineering issue can be solved by slightly shifting the radial position of the encoded phase at zone $P_8$, which does not compromise the effectiveness of the spatiotemporal data transmission.

**Discussion**

Beyond the finite 8-/9-bit encoding schemes demonstrated, our platform inherently supports scalable spatiotemporal coding with a theoretically unlimited number of bits by exploiting a pulse-stream expansion technique. Given that modern femtosecond lasers achieve repetition frequencies of 1–100 GHz [51-53], we assume a conservative 10-GHz repetition rate for our pulsed laser system. The transmitter core is a programmable phase modulator that simultaneously imposes spatial phase modulation and focuses the beam. Its architecture comprises 1000 zones (Fig. 6a), each with 100-fs temporal resolution, introducing a 100-ps propagation delay between pulses from the geometric center and those from the outermost zone ($P_{1000}$). This differential path length splits an input pulse into independently modulated sub-pulses with controllable temporal delays, which arrive at the focus in a temporally ordered sequence—thereby realizing spatial-to-temporal conversion. Each subsequent input pulse encounters a dynamically updated phase pattern, generating a new encoded pulse chain that maintains precise timing relative to its predecessor. Two critical temporal parameters govern this process: $\Delta T$=100 ps, dictated by the 10-GHz input repetition rate, defines the interval between successive pulse chains; and $\Delta t_{n \to n+1}$ =100 fs specifies the temporal interval between the end of one chain and the start of the next. Within each chain, pulses maintain uniform 100-fs spacing via precise zone modulation, allowing encoded information to be seamlessly concatenated across successive chains. Under steady-state laser operation, the system thus generates a continuous stream of non-overlapping, temporally ordered pulse sequences, enabling gapless information transmission. Simulations based on the phase modulator (NA = 0.447; see Supplementary Section 5) yield a single-channel data rate of 10 Tbit/s.

Building upon the continuous encoding architecture, a precise phase refresh mechanism is essential to ensure error-free operation. As illustrated in Fig. 6b, the phase pattern is pre-loaded before the first pulse arrives. When this pulse traverses the modulator, its sub-pulses are encoded according to this initial pattern. Immediately thereafter, the phase pattern updates sequentially, starting with the innermost zone ($P_1$) after 100 fs according to the next data bit. This radial refresh propagates outward ($P_1$–$P_{1000}$) at successive 100-fs intervals, completing a full cycle in 100 ps—in perfect synchrony with the arrival of the next input pulse. The cycle repeats with each pulse, ensuring that the modulator's 10-GHz refresh rate remains synchronized with the laser pulse train.

Although current commercial SLMs cannot operate at such frequencies, ongoing advances in electro-optic materials (e.g., lithium niobate and barium titanate) and fabrication techniques have already demonstrated GHz-level[54-56]. Further platform refinements are expected to render 10-GHz modulation feasible in the near future.

The multi-zone architecture, combined with the dynamic refresh mechanism, enables continuous, gapless data transmission. More significantly, it introduces a scalable paradigm for performance enhancement. Unlike static approaches, our platform offers programmable control over two key degrees of freedom: the number of sub-pulses and their individual durations. The latter allows the sub-pulse duration per zone to be actively compressed, directly increasing the data rate. This dual spatiotemporal programmability—absent from other existing space-time architectures—establishes a critical pathway toward radically scaled capacities in future optical communications.

In summary, we have experimentally demonstrated an all-optical transmitter architecture that encodes data into tailored spatiotemporal wave packets for single-channel transmission. By independently controlling the temporal width and number of coding zones, this scheme offers a scalable route to terabit-per-second throughput, bypassing the fundamental limits of electronic approaches. This work thus delivers both a pivotal solution for data-center capacity scaling and a versatile platform for optical field engineering, with implications spanning ultrafast laser processing, high-dimensional communications, and fundamental light-matter studies.

## Methods

### Experimental setup

The experimental setup used a femtosecond laser (central wavelength: 1.027 μm, pulse width: 200 fs, repetition rate: 60 Hz) as the input light source. In the signal arm, a motorized optical delay line (translation resolution: ±1 μm) provided programmable path-length adjustment. The beam was expanded by a 6× Galilean telescope ($L_1$: f = -50 mm, concave; $L_2$: f = 300 mm, convex). Phase modulation was imparted by a reflective phase-only spatial light modulator (SLM; HDSLM80R, 1920×1200 pixels, 8 μm×8 μm pixel pitch, 60 Hz refresh rate). The encoded phase patterns were then transferred to a phase-type PDL via a 4*f* imaging system ($L_3$ and $L_4$: *f* = 50 mm plano-convex lenses), with the SLM and PDL positioned at the system's input and output planes, respectively. The PDL, fabricated by electron-beam lithography on fused silica substrates, comprised 1,636 concentric zones and had a 7-mm focal length at the wavelength of 1.027 μm. All spatiotemporal signals were collected using a high-numerical-aperture objective (63× magnification, NA = 0.95). In the reference arm, chromatic dispersion was compensated by chirped mirrors and optical power was equalized by a tunable filter. The resulting reference beam was then combined with the signal beam to generate clear interference patterns. At the receiver, a scientific CMOS camera (CS505CU, Thorlabs) recorded the data. The camera, SLM, and optical delay line were synchronized and controlled by a single computer to coordinate data acquisition. The acquired data were subsequently processed offline.

### Theory of spatiotemporal wave packet propagation

Spatiotemporal wave packets are generated via modulation with a SLM and focusing with a Fresnel zone plate (PDL). The incident pulse, derived from a stable femtosecond laser, has approximately Gaussian spatial and spectral profiles. It can thus be expressed as:

$$A(x_0, y_0, \omega) = e^{-\left(\frac{\omega-\omega_0}{\Delta\omega}\right)^2} e^{-\left(\frac{x_0^2+y_0^2}{a^2}\right)} \tag{3}$$

where $x_0$ and $y_0$ are the transverse Cartesian coordinates on the input plane, $a$ is the full-width at half-maximum (FWHM) of the spatial profile, and $\omega$, $\omega_0$ and $\Delta\omega$ represent the angular frequency, central angular frequency, and angular frequency bandwidth of the pulse, respectively. The SLM imparts a phase modulation $U(x_0, y_0)$ onto the incident field. Therefore, the reflected field immediately after the SLM is:

$$E(x_0, y_0, \omega) = U(x_0, y_0) A(x_0, y_0, \omega) \tag{4}$$

Assuming an ideal 4f system (neglecting aberrations and edge diffraction), the field after the PDL can be expressed as:

$$T(x_0, y_0, \omega) = E(x_0, y_0, \omega)e^{i\phi(r_0)}, \quad r_0 = \sqrt{x_0^2 + y_0^2} \tag{5}$$

The phase profile of the PDL, $\phi(r_0)$, is given by:

$$\phi(r_0) = \begin{cases} 0, & d_{n-1} < r_0 \leq d_n \\ \pi, & d_n < r_0 \leq d_{n+1} \end{cases} \quad n = 1, 2, 3, \cdots \tag{6a}$$

$$d_n = \sqrt{n\lambda_0 f + \frac{1}{4}n^2\lambda_0^2} \tag{6b}$$

where $n$ and $f$ denote the zone index and focal length of the PDL, respectively, and $\lambda_0$ is the central wavelength of the laser pulse.

According to the Rayleigh-Sommerfeld diffraction integral[49,50], the optical field at a propagation distance $z$ can be expressed as:

$$E(x, y, z, \omega) = \frac{\partial}{\partial z} \iint_S T(x_0, y_0, \omega) \cdot \frac{e^{ikR}}{R} dx_0 dy_0 \tag{7}$$

where $k = \omega/c$, $S$ denotes the integration region on the input plane, and $R = \sqrt{(x-x_0)^2 + (y-y_0)^2 + z^2}$. The temporal field distribution $E(x, y, z, t)$ is obtained by taking the inverse Fourier transform of $E(x, y, z, \omega)$, which yields (up to a constant factor):

$$E(x, y, z, t) = \int_{-\infty}^{\infty} E(x, y, z, \omega) e^{-i\omega t} d\omega \tag{8}$$

By combining Eq. 3-5, 7 and 8, $E(x, y, z, t)$ can be rewritten as:

$$E(x, y, z, t) = \frac{\partial}{\partial z} \iint_S \left[ \int_{-\infty}^{\infty} U(x_0, y_0) e^{-\left(\frac{\omega-\omega_0}{\Delta\omega}\right)^2} e^{-\left(\frac{x_0^2+y_0^2}{a^2}\right)} e^{i\phi(r_0)} \frac{e^{ikR}}{R} e^{-i\omega t} d\omega \right] dx_0 dy_0 \tag{9}$$

In polar coordinates, the output field is:

$$E(r, \theta, z, t) = \frac{\partial}{\partial z} \int_{r_{in}}^{r_{out}} \int_0^{2\pi} \left[ \int_{-\infty}^{\infty} U(r_0, \theta_0) e^{-\left(\frac{\omega-\omega_0}{\Delta\omega}\right)^2} e^{-\left(\frac{r_0^2}{a^2}\right)} e^{i\phi(r_0)} \frac{e^{ikR}}{R} e^{-i\omega t} d\omega \right] r_0 dr_0 d\theta_0 \tag{10}$$

where $r_{in}$ and $r_{out}$ are the inner and outer radii (as defined in Eq. (2)) of a given annular encoding zone, and $\theta_0$ is the azimuthal angle on the input plane. The corresponding expression for $R$ is given by $R = \sqrt{r_0^2 + r^2 + z^2 - 2r_0 r \cos(\theta_0 - \theta)}$. The spatiotemporal distribution at the focal plane of the PDL ($z = f$) is obtained by sweeping the angular frequency and summing the results.

Numerical simulations were performed using the theory described above. The input plane had a pixel size of 0.5 μm. The central wavelength was 1.027 μm, and the wavelength range was sampled from 1.027 μm ± 10 nm with a step size of 1 nm. To enhance the spatial resolution at the output plane, the light field was first propagated to an intermediate plane at $z = 0.8\,f$. The resulting complex field was then replicated and up-sampled by a factor of 5. Finally, the field at the output plane ($z = f$) was obtained by propagating this densified field over the remaining distance of $0.2f$, yielding an output pixel size of 0.1 μm.

## Code availability

All code used in this study is available from the corresponding authors upon request.

## Acknowledgement

The authors thank the National Natural Science Foundation of China (Grant Nos. 62322512, 62225506, 62505308 and 12134013), the Fundamental Research Funds for the Central Universities (WK2030000108 and WK2030000090), CAS Project for Young Scientists in Basic Research (Grant No.YSBR-049) and the support from the USTC Center for Micro and Nanoscale Research and Fabrication. D.Z. thanks the China Postdoctoral Science Foundation under Grant Number 2023M743364 and Anhui Natural Science Foundation (2508085QA010).

## Author contributions

K. H. conceived the idea. W.-J. L., H.-Y. W. and J. H. carried out the simulations. D. Z. and H.-Y. W. fabricated the sample. W.-J. L. and H.-Y. W. built up the experiment setup and carried out the measurement. F.-W. S. and Y. T. contributed to data analysis. W.-J. L., K. H., F.-W. S. and Y. T. wrote the manuscript with inputs from other authors. K. H. and Y. T. supervised the overall project. All authors commented on this manuscript.

## Competing financial interests

The authors claim no competing interests.

# Figures and Captions

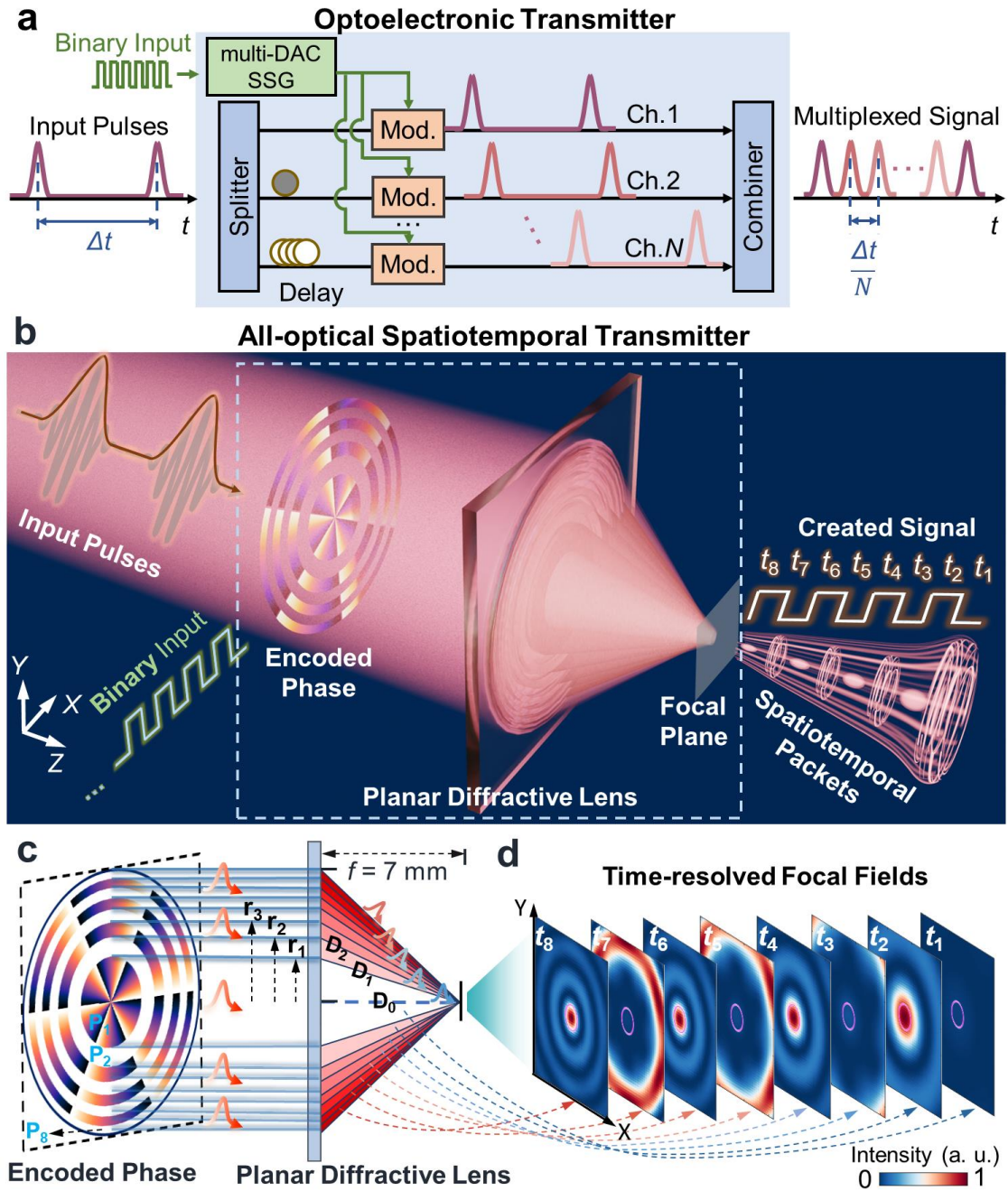

**Figure 1 | Traditional optoelectronic and all-optical spatiotemporal transmitters. a,** Principle of traditional optoelectronic transmitter. An input pulse train (interval $\Delta t$) is routed into an optoelectronic transmitter and split into $N$ parallel branches. Each branch is independently modulated by an electrical drive signal from the multi-DAC synchronous signal generator (SSG), controlled by the binary input, and delayed by a predefined temporal offset. The branches are subsequently recombined to form a temporally interleaved pulse train (interval $\Delta t/N$). Multi-DAC SSG, comprising a serial-to-parallel converter, multiple DACs, and drivers. Mod., electro-optic modulators. Ch. 1 to Ch. $N$ denote

Channels 1 to *N*. **b,** Sketch for all-optical spatiotemporal transmitter. The input pulses are modulated by an integrated all-optical spatiotemporal transmitter comprising an encoded phase pattern and a PDL, generating STWPs for data transmission. **c,** Detailed workflow of the phase-encoding methodology. An 8-bit binary data sequence is mapped onto eight annular zones, each encoded with a phase aligned to the PDL. **d,** Simulated focal-plane intensity distributions normalized to the peak intensity at each time slot. Profiles taken at the mid-temporal position of each annular zone show the energy variation across the STWP. The magenta circles indicate the detection regions.

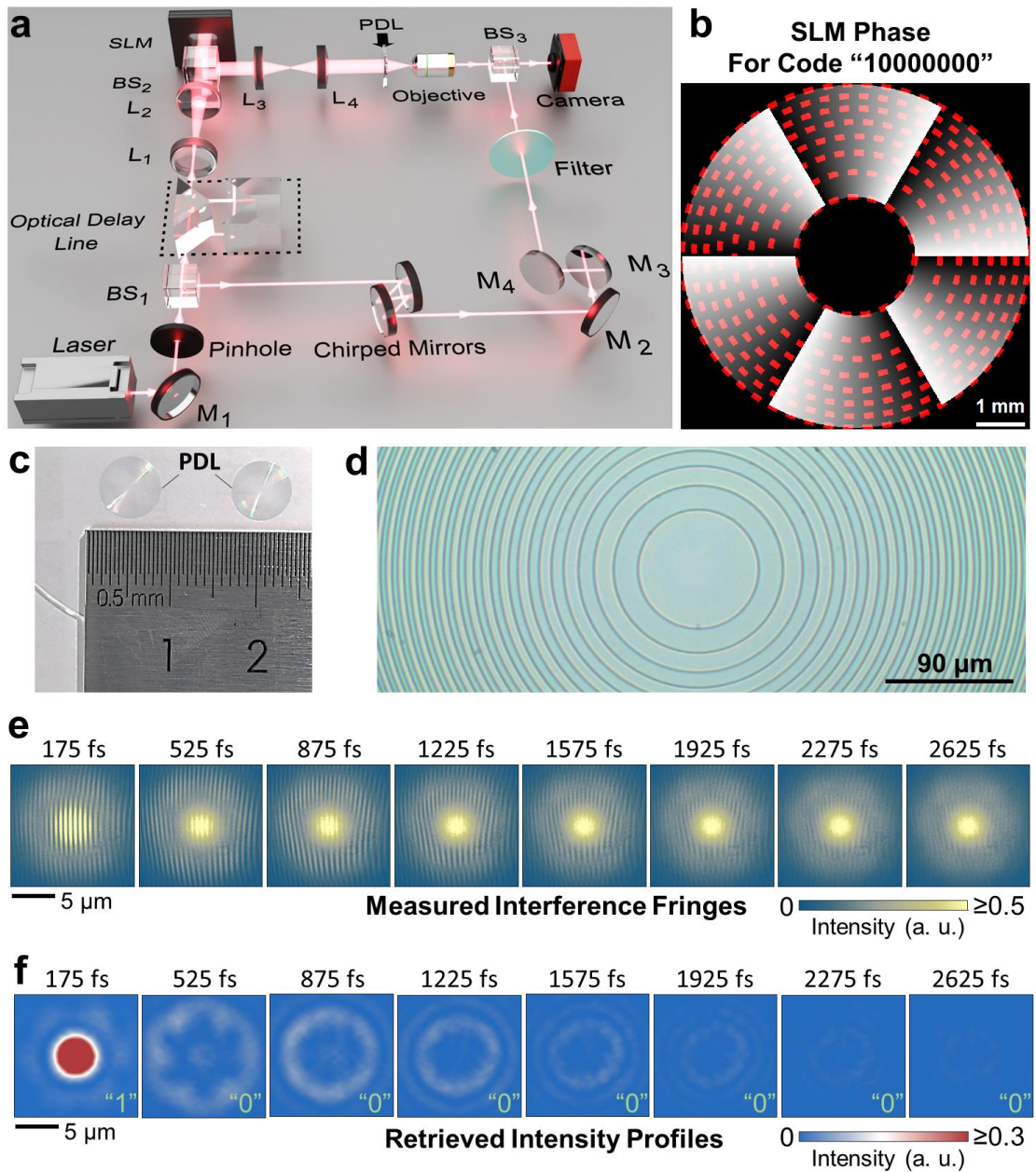

**Figure 2 | Generating and detecting data-carrying spatiotemporal wave packets. (a)** Experimental setup for wave-packet characterization. $M_1$–$M_4$: mirrors; $BS_1$–$BS_3$: beam splitters; $L_1$–$L_4$: lenses; SLM:

spatial light modulator; PDL: planar diffractive lens. **(b)** Phase pattern loaded onto the SLM encoding an exemplified code "10000000". Red circles delineate the eight zones. **(c)** Image of our fabricated FDL. **(d)** Magnified view of the center of the fabricated PDL, imaged with a 10× objective. **(e)** Interference fringes recorded by the camera at eight time instants. Intensity is normalized to the global maximum. For improved visibility of weak-intensity fringes, we apply pseudo-coloring only to intensity values between 0 and 0.5. **(f)** Signal intensity profiles extracted from the fringes in **(e)**, normalized to the global maximum. Similarly, we apply pseudo-coloring only the normalized intensity values between 0 and 0.3.

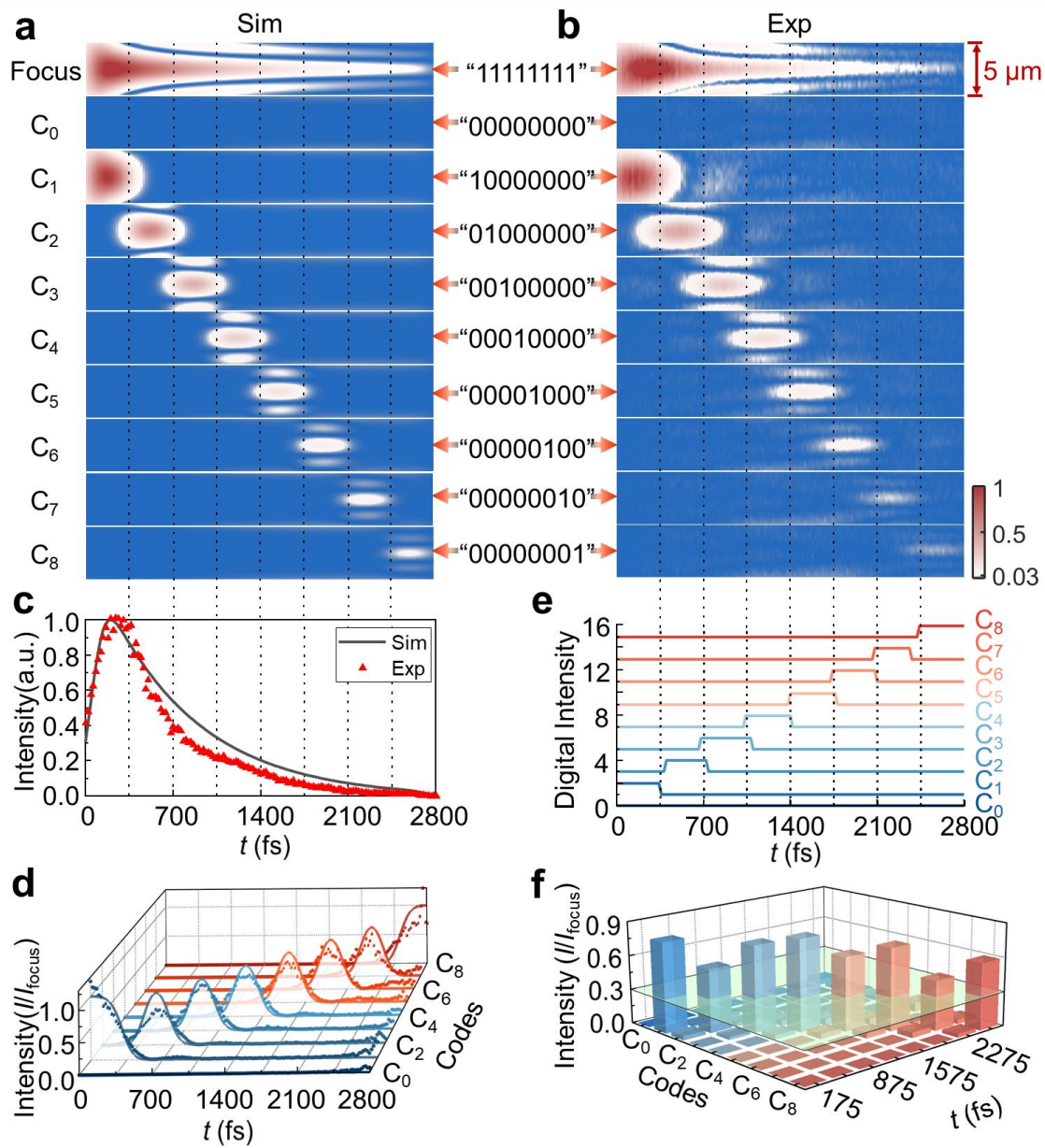

**Figure 3 | Inter-bit temporal orthogonality. (a-b)** Simulated **(a)** and **(b)** experimental intensity

profiles on the *x-t* plane within the focal region for different codes. Corresponding codes are indicated centrally. **(c)** Temporal evolution of the normalized total intensity for the all-"1" state ("11111111"), integrated within a circular region in the focal plane (*x-y* plane). The region diameter equals the focal spot's full width at half maximum (FWHM) at 1.96 ps. **(d)** Ratio of the total intensity for each code $C_0$–$C_8$ to that of all-"1" (focused) state ($I/I_{focus}$). Solid curves and dots denote simulated and experimental data, respectively. **(e)** Digitalized binary temporal sequences decoded by applying a threshold of 0.3 to the experimental intensity profiles in **(d)**. **(f)** Experimental normalized intensities of $C_0$–$C_8$ sampled at the midpoint of each 350-fs time window in **(e)**. The clear separation of all signals from the 0.3-threshold plane (light green) confirms the retention of temporal orthogonality.

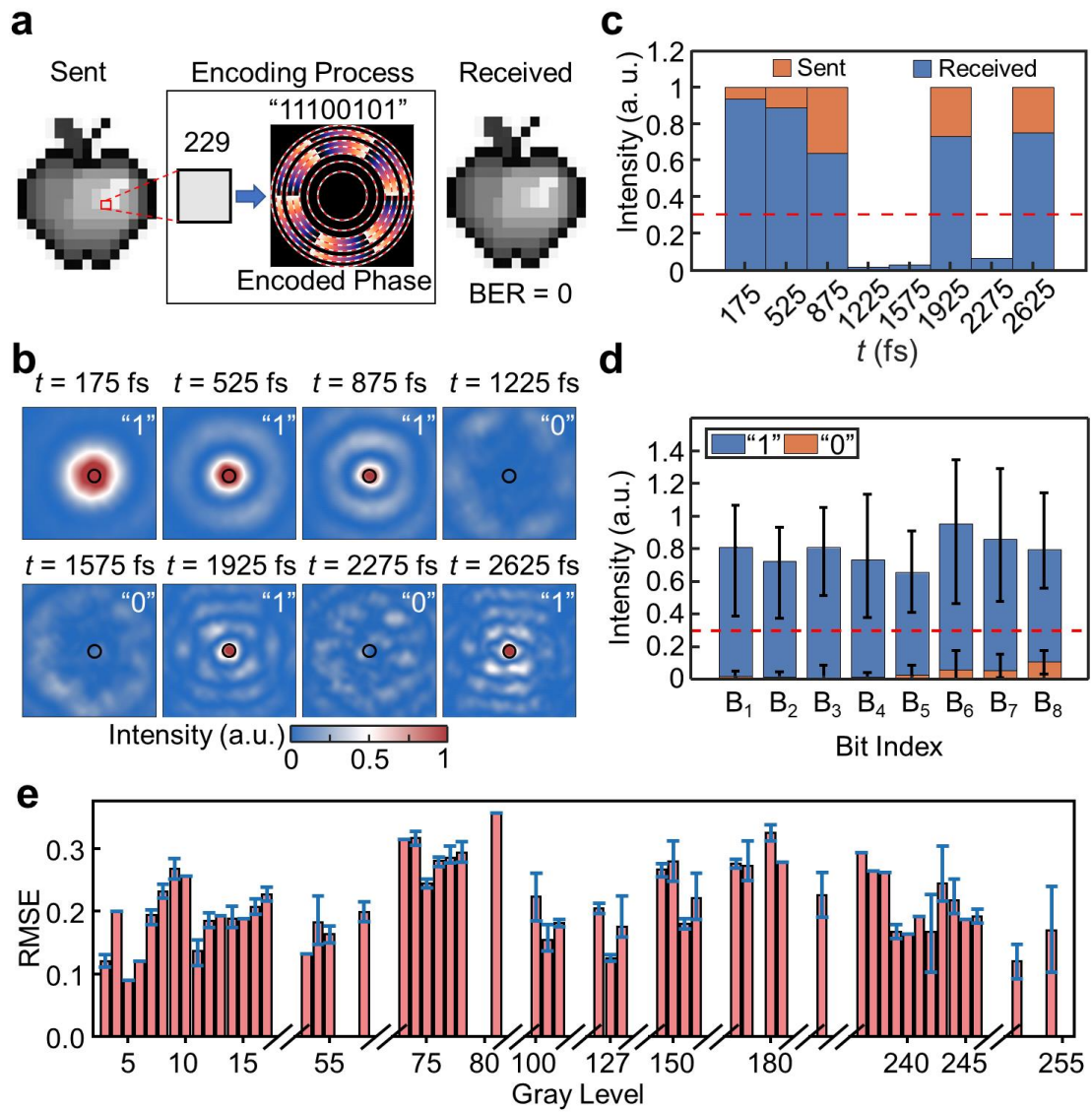

**Figure 4 | Grayscale image transmission via 8-bit STWPs encoding. (a)** Image transmission processes including sent image (left), encoding process (middle) and received image (right). Each gray

level in the original image is mapped to an 8-bit binary sequence and encoded onto the STWP. Reconstructed grayscale image is obtained from the experimentally retrieved data. **(b)** Intensity distributions (*x-y* plane) at eight temporal sampling nodes within the focal region for a representative pixel (marked in **(a)**). Black circles outline the regions where the intensity is integrated. **(c)** Comparison of normalized experimental intensities within the detection region (blue bars) with the ideal digital values (orange bars). The red dashed line marks the threshold (0.3) for binary-state discrimination. **(d)** Experimentally normalized intensities for logic "1" (blue) and "0" (orange) across all bits of the transmitted image in **(a)**. Error bars span the minima to maxima of the measured values. **(e)** Root-mean-square errors (RMSEs) between the experimental and digital intensities for all gray levels in the transmitted image.

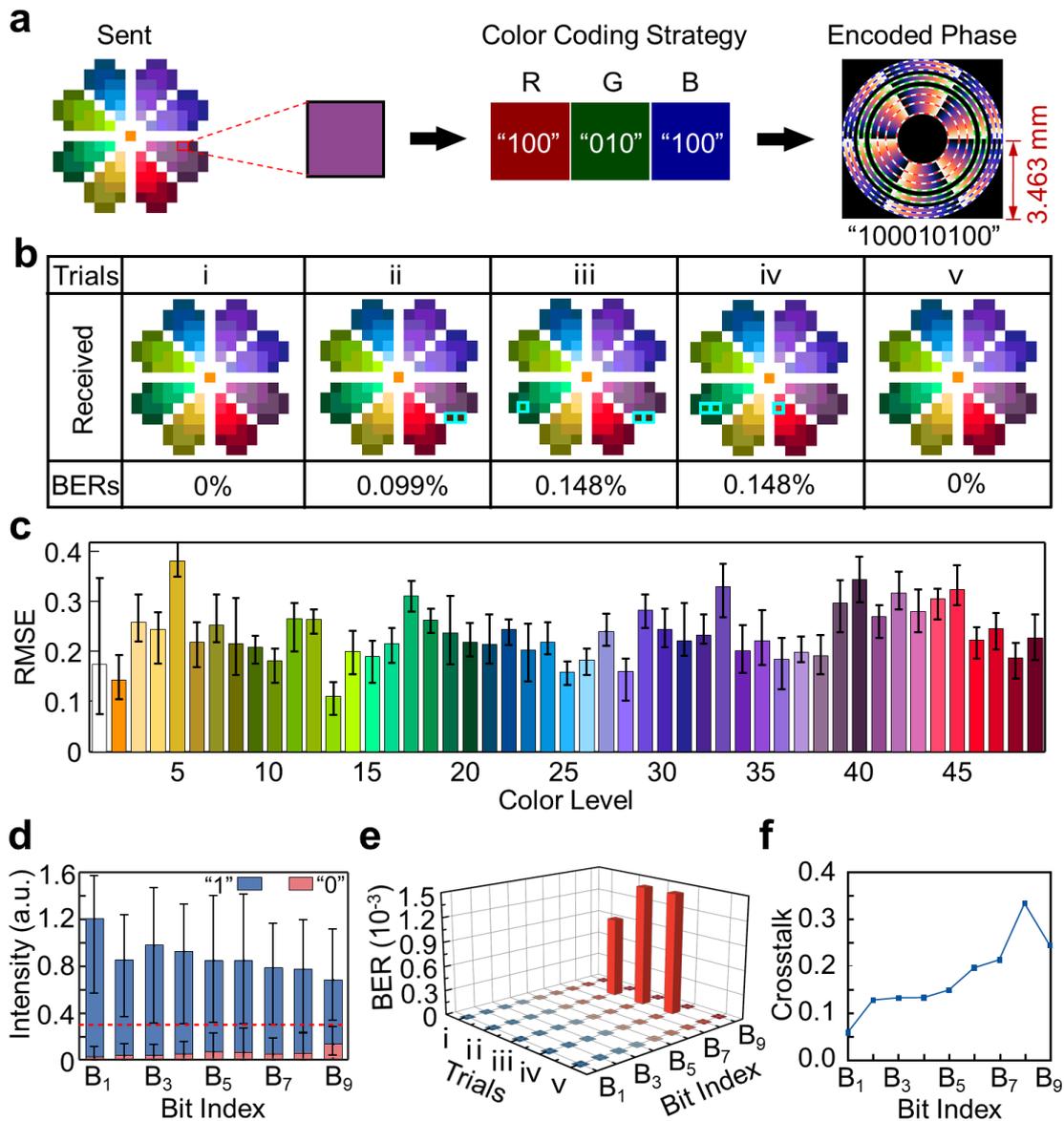

**Figure 5 | Full-color image transmission employing 9-bit data encoding.** (a) Schematic of encoding a RGB color into a 9-bit data via 9 zones with different phase profiles. Zones $P_1$–$P_9$ (from the center to the outmost) correspond to bit positions 1–9, respectively, arranged from left to right. Dashed circles (red, green, and blue) delineate the boundaries of the annular zones corresponding to each color channel. (b) Received images from five transmission trials, with erroneous pixels highlighted (light-blue boxes), and their individual BERs. (c) RMSEs for all color levels. Error bars span the range (minima to maxima) across trials. (d) Normalized intensity distributions for logic "1" (blue bars) and "0" (pink bars) at each bit position during transmitting this color image. Error bars denote experimental variability (five trials). (e) BERs for each bit position ($B_1$-$B_9$) aggregated from five trials. (f) Inter-logic-state crosstalk per bit position.

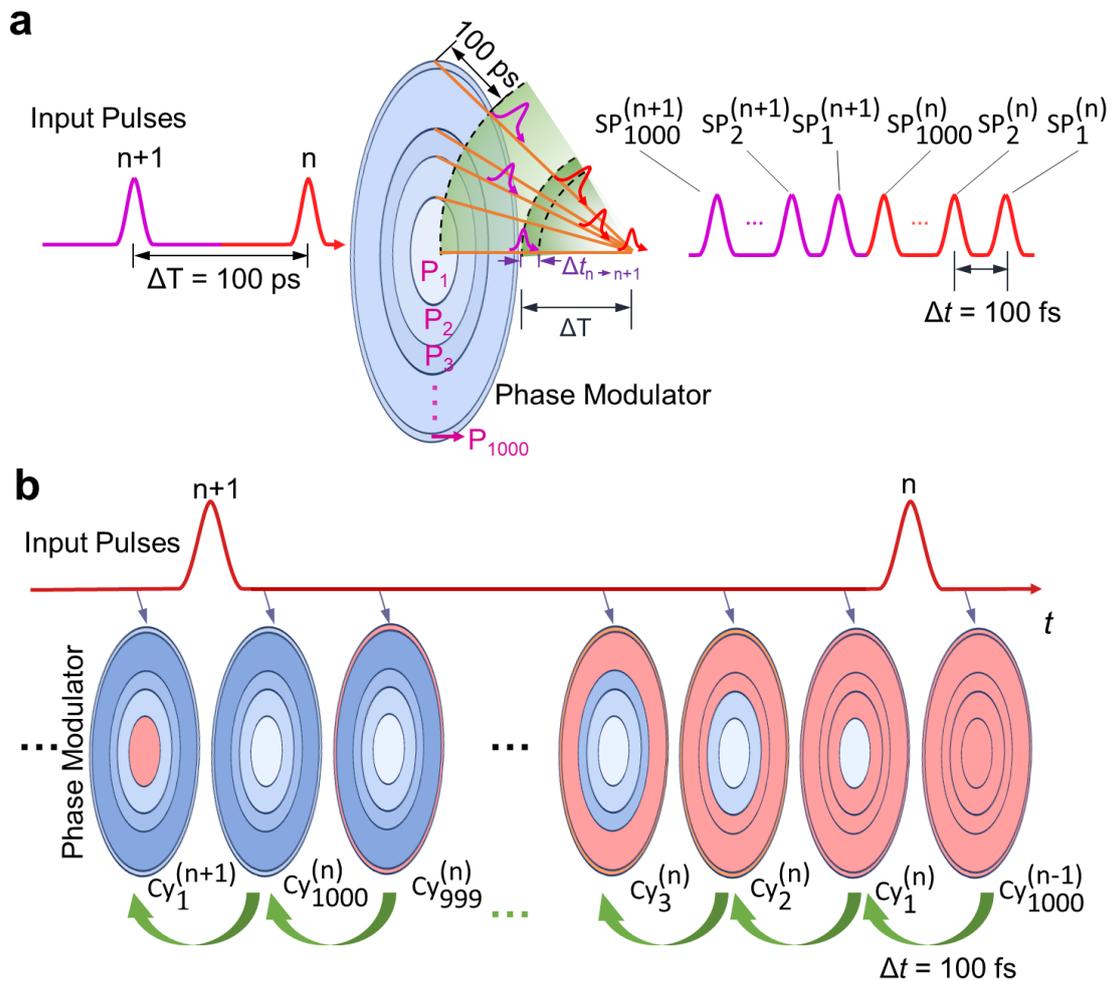

**Figure 6 | Proposal for continuous-bit spatiotemporal coding.** (a) Schematic of the spatial-to-temporal pulse expansion method. Low-repetition-rate input pulses are modulated by a phase modulator to generate a contiguous high-repetition-rate pulse train that carries encoded information.

The temporal interval between the last sub-pulse of the $n^{th}$ incident pulse and the first sub-pulse of the $(n+1)^{th}$ incident pulse is $\Delta t_{n \to n+1} = 100$ fs. For clarity, the SLM phase and the PDL phase are integrated into the phase modulator; only the zones corresponding to the SLM phase are highlighted here. **(b)** Dynamic phase-refresh mechanism of the low-rate modulator that ensures error-free, continuous data transmission. The modulator states are denoted as $Cy_m^{(n)}$, where $n$ is the index of the incident pulse (cycle number) and m is the index of the sub-pulse generated from that incident pulse (iteration number within the cycle). In our scheme, the modulator is updated at a low rate of 100 ps —as illustrated by the change of the centre circles from $Cy_1^{(n)}$ to $Cy_1^{(n+1)}$ —whereas subpulses are generated at a time interval of 100 fs, which significantly reduce the requirement for SLM's refreshing rate. The psudeo-colors denote changes in the modulation and do not represent specific phase values.

Supplementary Materials for

# Spatiotemporal flat optics for terabit-per-second single-channel data transmission

Wen-Jing Liu[1,#], Dong Zhao[1,#,*], Heng-Yi Wang[1,#], Jun He[1], Fang-Wen Sun[1], Ye Tian[2,3,*], Kun Huang[1,4,*]

[1]Department of Optics and Optical Engineering, University of Science and Technology of China, Hefei, 230026, Anhui, China

[2]State Key Laboratory of Ultra-intense Laser Science and Technology, Shanghai Institute of Optics and Fine Mechanics (SIOM), Chinese Academy of Sciences (CAS), Shanghai 201800, China

[3]Center of Materials Science and Optoelectronics Engineering, University of Chinese Academy of Sciences, Beijing 100049, China

[4]State Key Laboratory of Opto-Electronic Information Acquisition and Protection Technology, Anhui University, Hefei, 230601, Anhui, China

**Table of Contents**

# Supplementary Section 1 | Spatiotemporal field's retrieval from interference fringes

All experimental data were recorded as interference patterns for signal field retrieval. Supplementary Fig. 1 illustrates the processing procedure for one such pattern recorded at $t = 175$ fs. At this time delay, a nearly vertical interference fringe was observed (Supplementary Fig. 1a). Its spatial spectrum, obtained via a two-dimensional Fourier transform, is shown in Supplementary Fig. 1b. A bandpass filter was used to isolate the +1st (or -1$^{st}$) diffraction order in the spatial-frequency domain. This region contains the complex data—the amplitude (Supplementary Fig. 1b, i) and phase (Supplementary Fig. 1b, ii) information of the signal. An inverse Fourier transform of this filtered component then reconstructed the signal field. The resulting intensity distribution (Supplementary Fig. 1c) exhibits a focused spot profile. The same processing procedure was applied to interference patterns recorded at other time delays (Supplementary Fig. 1d), yielding the corresponding signal-field intensity distributions shown in Supplementary Fig. 1e. As the time delay increases, both the interference fringes and the corresponding retrieved signal-field distributions evolve. The signal field forms a focal spot at time delays 175 fs to 1225 fs, 2275 fs and 2625 fs, whereas it exhibits a ring-shaped distribution at 1575 fs and 1925 fs.

# Supplementary Section 2 | System signal-to-noise ratio

The signal-to-noise ratio (SNR) of the system was evaluated using experimental data from the 8-bit transmission measurement shown in Fig. 3. The signal $I_{focus}$ is defined as the detected intensity for the all-"1" encoding, and the noise floor $I_0$ as the intensity for the all-"0" encoding. The resulting SNR ($I_{focus}/I_0$) is shown in Supplementary Fig. 2. A gradual temporal decay in SNR is observed, which correlates with the signal-intensity decay in Fig. 3c. Nevertheless, the SNR remains above 10 dB throughout the measurement window. These values represent a conservative estimate of system performance, as a portion of the signal was attenuated by an infrared filter for camera protection. A higher SNR is therefore attainable in practice.

# Supplementary Section 3 | 9-bit orthogonal-matrix transmission

To assess the system's performance at a higher data rate, we conducted a 9-bit orthogonal-matrix transmission experiment. Here, the duration per encoding zone was reduced from 350 fs to 300 fs, shortening the bit period accordingly. Supplementary Fig. 3a displays the temporally resolved intensity distribution within the focal region under different encoding schemes, covering the 2.7-ps window corresponding to nine bits. The all-"1" sequence ("111111111") again serves as the reference. Its temporal profile matches that in Fig. 3c, confirming a consistent focusing state. Following the same encoding principle, the all-"0" code ($C_0$) produces no focal spot, while any code containing a "1" ($C_1$–$C_9$) generates a spot whose arrival time shifts linearly with the bit position of the "1". For the 9-bit configuration, this shift is 300 fs per bit, which—given the reduced bit period—corresponds to a proportionally higher data rate. The corresponding experimental measurements (Supplementary Fig. 3b) confirm this behavior and demonstrate that the system reliably generates the predicted spatiotemporal pattern even at the shorter 300 fs bit period.

For quantitative analysis, the same detection region as defined in the main text was used, based on the FWHM of the reference focal spot at $t = 1.96$ ps. The intensity profile of the all-"1" reference (Supplementary Fig. 3c) matches that in Fig. 3c, confirming a stable focusing state. Supplementary Fig. 3d displays the corresponding traces for all codes, showing clear temporal peaks only for codes containing a "1" ($C_1$–$C_9$). The linear shift of these peaks yields an interval of 300 fs per bit—shorter than the 350 fs in the 8-bit scheme—corresponding to a data rate of 3.33 Tbit s$^{-1}$. Applying an intensity threshold converts these traces into digital signals (Supplementary Fig. 3e), where a single high-logic level propagates sequentially through the nine 300-fs time windows. Following the discrete-sampling approach established in the main text, we sampled the intensity near the midpoint of each 300-fs window. The normalized intensities at these nine points (Supplementary Fig. 3f) confirm that the orthogonal encoding retains its distinguishability even at this higher data rate: only the time slot containing a "1" exceeds the threshold. Thus, the 9-bit input code can be read directly from this pattern of intensity states.

## Supplementary Section 4 | Experimental demonstration of a 9-bit sequence

Supplementary Fig. 4 shows the experimental transfer of a 9-bit coded sequence ("100010110"). Because the 2$^{nd}$, 3$^{rd}$, 4$^{th}$, 6$^{th}$, and 9$^{th}$ bits are "0", the corresponding zones exhibit vortex-phase distributions (Supplementary Fig. 4a). The measured interference fringes at nine time instants become progressively blurred with increasing time delay (Supplementary Fig. 4b). As shown in Supplementary Fig. 4c, at a delay of 150 fs, the retrieved signal appears as a focused spot. From 450 fs to 1050 fs, the light field evolves into a donut-shaped pattern, with weak residual central intensity attributed to inter-symbol crosstalk. This residual arises from slight temporal misalignment of the sampling instant relative to the centre of the temporal window. At 1350 fs, 1950 fs, and 2250 fs, distinct focal spots are observed, consistent with the input bits. The normalized total intensity $I_{total}$ exceeds 0.3 for bit "1" and remains below this level for bit "0". The decoded sequence matches the input, confirming stable operation upon scaling from 8 to 9 bits.

## Supplementary Section 5 | Scalability of the spatiotemporal encoding architecture

To numerically assess the scalability of the system, we simulated its performance under representative encoding patterns as the number of zones was increased to 1000 (Supplementary Fig. 5). The phase modulator was designed with a numerical aperture (NA) of 0.4472. Using Eq. 2 in main text, we calculated the corresponding focal length and outermost zone radius to be $f = 254.16$ mm and $r_0 = 127.08$ mm, respectively (see Supplementary Fig. 5a). The inner and outer radii of each zone were derived from the same relation.

We first simulated the transmission of an all-"1" sequence by applying a focusing phase to all zones. The resulting spatiotemporal packets (STWPs) were calculated via Eq. (10), with $\phi(r_0)$ given by the non-paraxial focusing profile $\phi(r_0) = -k(\sqrt{r_0^2 + f^2} - f)$. The output plane was discretized with a pixel size of 0.1 μm, and the field was evaluated at the temporal midpoint of each zone. To reduce the computational burden, the integration limits $r_1$ and $r_2$ in Eq. 10 were restricted to the radial extent of the corresponding zone when calculating the intensity at each moment. The intensity distribution on the *x-t* plane (Supplementary Fig. 5b) exhibits a sustained focal state across all 1000 time instants.

We next simulated an alternating "1" and "0" sequence by imprinting a vortex phase (topological charge 6) onto the even-numbered zones. The corresponding intensity profile (Supplementary Fig. 5c) reveals a periodic absence of energy at even time slots. The total intensity within a central circular detection region (radius 1 μm) was normalized for both cases. For the all-"1" case, the detected intensity gradually decays (Supplementary Fig. 5d). In the alternating sequence, the intensity drops to near zero at every even moment but remains nearly constant at odd moments (Supplementary Fig. 5e). Notably, the ratio of the intensities in Supplementary Fig. 5e to those in Supplementary Fig. 5d recovers the original alternating "1" and "0" bit stream (Supplementary Fig. 5f). With a 100-fs interval between adjacent bits, this scheme achieves a data rate of 10 Tbit/s.

These simulations confirm that the architecture maintains error-free operation and the target data rate even when scaled to 1000 zones, demonstrating its scalability for high-capacity transmission.

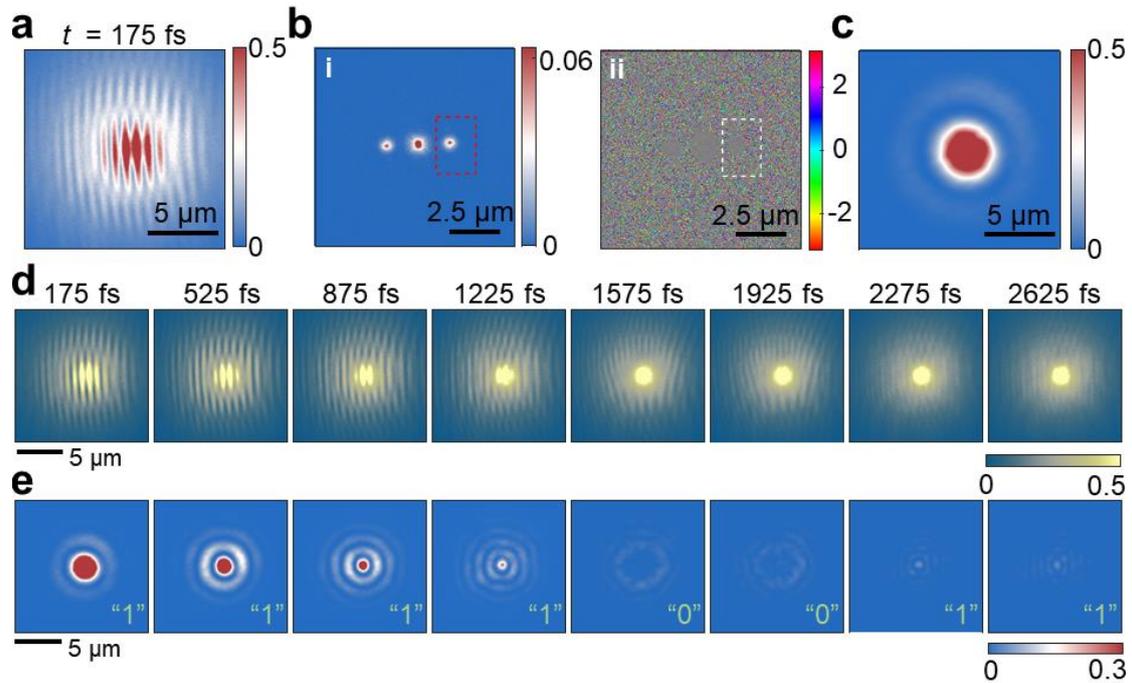

**Supplementary Figure 1 | Signal extraction from interference patterns.** (**a**) Experimentally recorded interference pattern on the *x–y* plane for code "11110011" at *t* = 175 fs (normalized to its maximum intensity). (**b**) Partial spatial-frequency spectrum obtained by two-dimensional Fourier transform of the fringe. The red box outlines the region of the +1st diffraction order; the complex spectral data (amplitude, **i**; phase, **ii**) within this region are used for the inverse Fourier transform. (**c**) Reconstructed signal-field intensity distribution after inverse Fourier transformation (normalized). (**d**) All experimental interference fringes for code "11110011". (**e**) Retrieved signal distributions from (**d**); the numbers at the bottom right indicate the decoded bits.

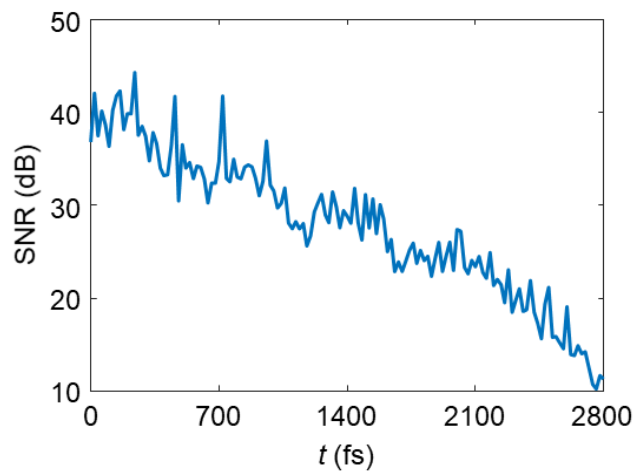

**Supplementary Figure 2 | Temporal evolution of the system's signal-to-noise ratio.** Signal-to-noise ratio (SNR) calculated from the 8-bit transmission experiment in Fig. 3c (main text).

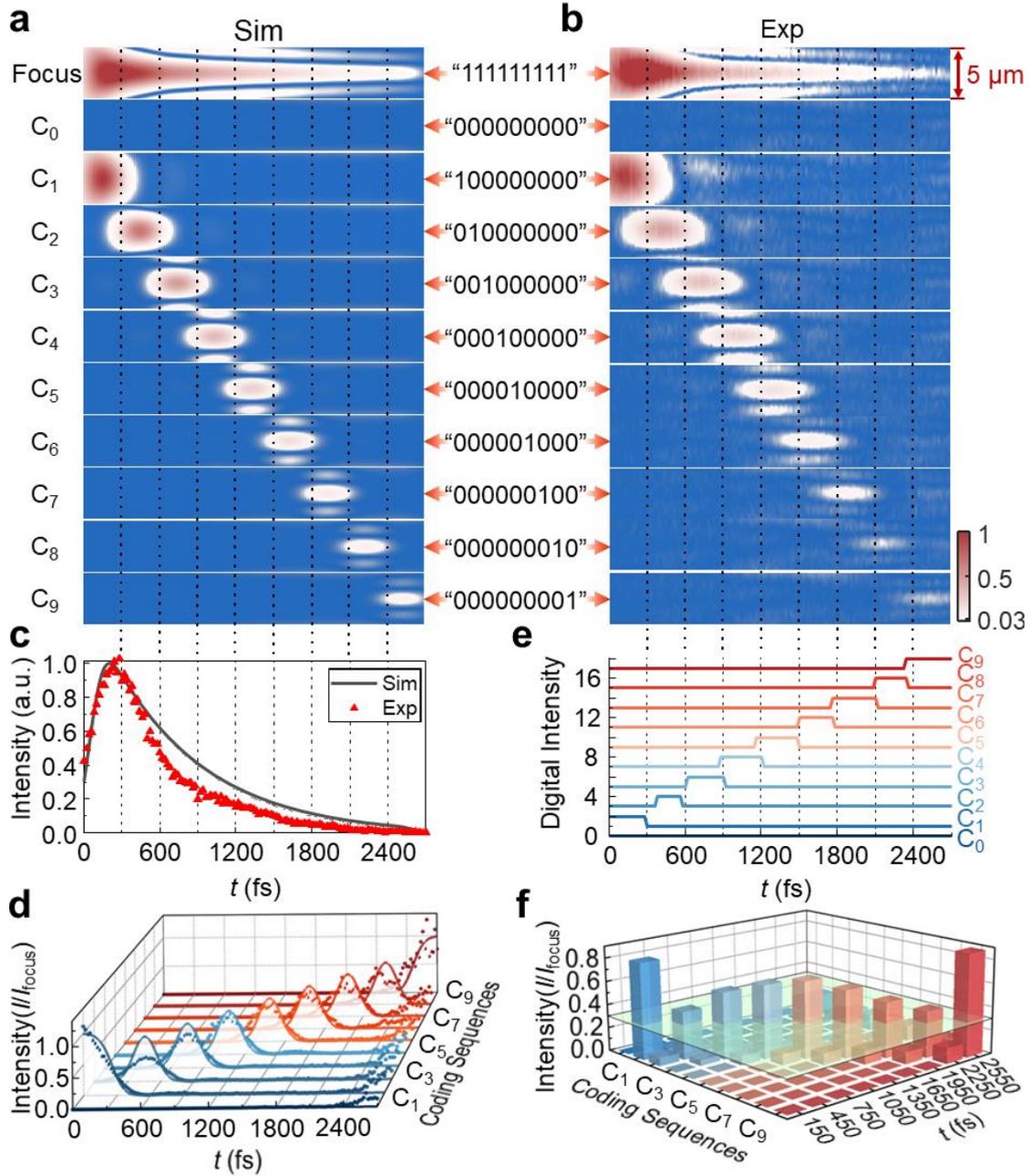

**Supplementary Figure 3 | 9-bit temporal orthogonality. (a-b)** Simulated (**a**) and experimental (**b**) intensity distributions on the *x-t* plane within the focal region for the indicated coding sequences. (**c**) Normalized total intensity within the detection region for the all-"1" coding sequence. (**d**) Normalized total intensity over time within the detection region for all coding sequences, where $I$ and $I_{focus}$ represent the intensity of the encoded sequence and the focused state, respectively. Solid and dashed lines denote simulated and experimental data. (**e**) Decoded binary sequences obtained by applying a threshold of 0.3 to the experimental profiles in (**d**). (**f**) Normalized intensity sampled at nine specific time instants.

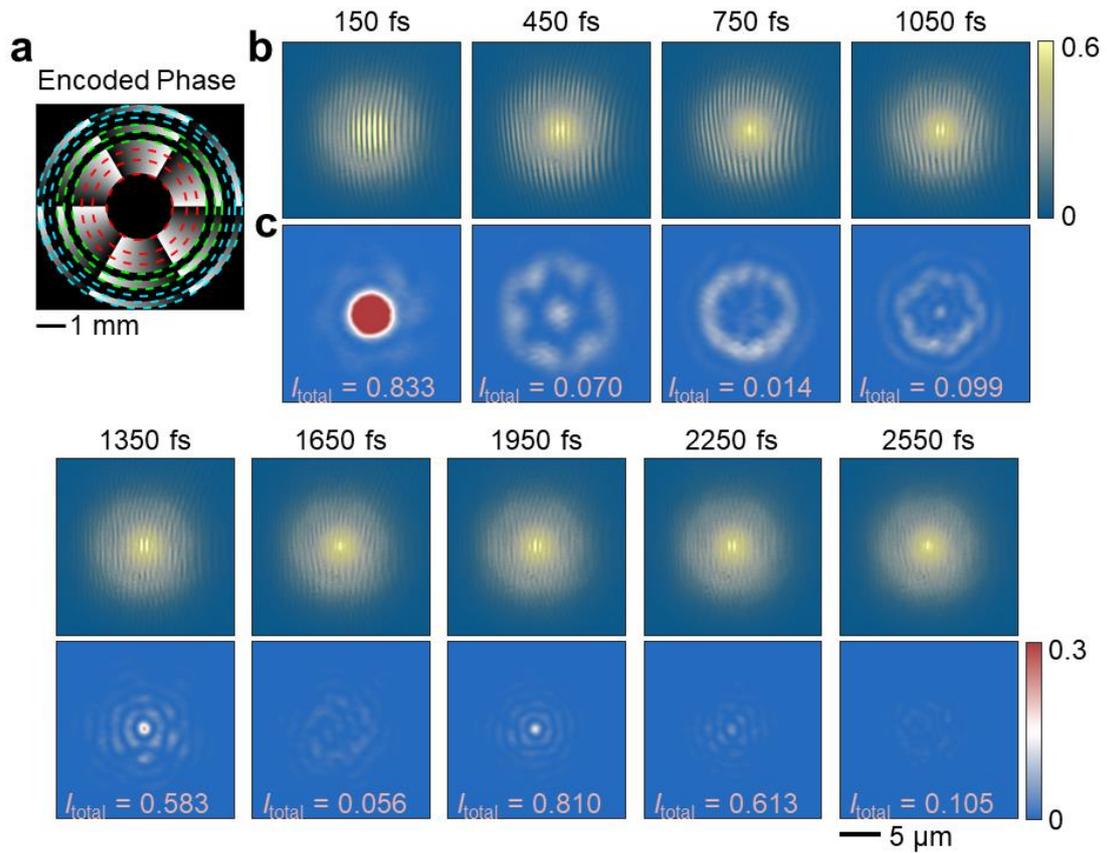

**Supplementary Figure 4 | Experimental results for the 9-bit sequence "100010110". (a)** Encoded phase for the bit sequence. Red, green and blue dotted lines mark the boundaries of the zones corresponding to the R, G and B channels, respectively. **(b)** Experimentally recorded interference patterns on the $x$–$y$ plane at nine time instants. Intensities are normalized to the global maximum. **(c)** Retrieved signal intensity distributions from **(b)**. For each distribution, $I_{total}$ at the bottom denotes the total intensity within the detection region, normalized to that of the all-focus condition.

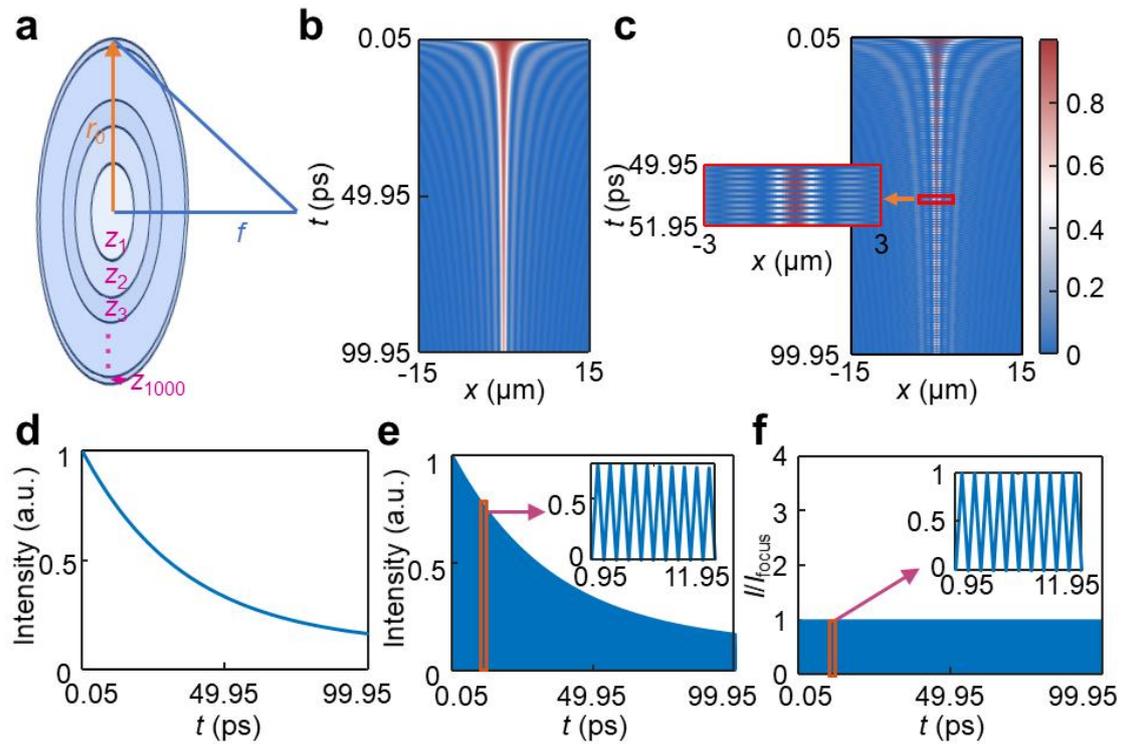

**Supplementary Figure 5 | Simulated transmission performance of a phase modulator with 1000 zones. (a)** Schematic of the phase modulator with 1000 zones. $r_0$ and $f$ denote the outermost radius and the focal length, respectively. **(b-c)** Normalized intensity distributions on the *x-t* plane for **(b)** all-"1" encoding and **(c)** alternating "1" and "0" encoding. In **(c)**, the red box magnifies the 41.95–51.95 ps region, revealing alternating intensity dropouts at even time slots. **(d-e)** Normalized total intensity (to the global maximum) within a circular detection region (radius: 1 μm) at the center of the focal plane over time for **(d)** all-"1" case and **(e)** the alternating-encoding case. The orange box in **(e)** highlights the time interval from 0.95 ps to 11.95 ps, detailed in the upper-right inset, demonstrating the drop in intensity to near zero at all even time slots. **(f)** Ratio of the encoded curve **(e)** to the focused curve **(d)**, showing alternating 1/0 levels that match the input binary sequence. The orange box shows a magnified view of this alternating 1/0 pattern within the 0.95–11.95 ps interval.